# Suspended graphene membranes with attached silicon proof masses as piezoresistive NEMS accelerometers

Xuge Fan, *,† Fredrik Forsberg, ∥ Anderson D. Smith, § Stephan Schröder,†

Stefan Wagner, ‡ Mikael Östling, § Max C. Lemme,*,‡,§ and Frank Niklaus*, †

†Department of Micro and Nanosystems, School of Electrical Engineering and Computer Science, KTH Royal Institute of Technology, SE-10044 Stockholm, Sweden.

‡ Faculty of Electrical Engineering and Information Technology, RWTH Aachen University, Otto-Blumenthal-Str. 25, 52074 Aachen, Germany.

§Department of Integrated Devices and Circuits, School of Electrical Engineering and Computer Science, KTH Royal Institute of Technology, SE-164 40 Kista, Sweden.

∥Scania Tekniskt Centrum, 15148 Södertälje, Sweden.

*email: xuge@eecs.kth.se, lemme@amo.de, frank.niklaus@eecs.kth.se

**ABSTRACT**

Graphene is an atomically thin material that features unique electrical and mechanical properties, which makes it an extremely promising material for future nanoelectromechanical systems (NEMS). Recently, basic NEMS accelerometer functionality has been demonstrated by utilizing piezoresistive graphene ribbons with suspended silicon proof masses. However, the proposed graphene ribbons have limitations regarding mechanical robustness, manufacturing yield and the maximum measurement current that can be applied across the ribbons. Here, we report on suspended graphene membranes that are fully-clamped at their circumference and that have attached silicon proof masses. We demonstrate their utility as piezoresistive NEMS accelerometers and they are found to be more robust, have longer life span and higher manufacturing yield, can withstand higher measurement currents and are able to suspend larger silicon proof masses, as compared to the previously graphene ribbon devices. These findings are an important step towards bringing ultra-miniaturized piezoresistive graphene NEMS closer towards deployment in emerging applications such as in wearable electronics, biomedical implants and internet of things (IoT) devices.

The monitoring of acceleration is essential in a broad range of applications such as navigation systems, automotive crash detection systems and structural monitoring. Typical microelectromechanical system (MEMS) accelerometers occupy die areas of the order of several square millimetres. Further miniaturization of MEMS accelerometers results in smaller components and packages, and ultimately in reduced costs, which is critical for emerging applications such as wearable electronics,[1] biomedical implants,[2] nanoscale robotics,[3] and the Internet of Things (IoT).[4] However, downscaling of MEMS accelerometers comprises miniaturization of the electromechanical transducer and the size of the proof mass, thereby severely reducing device sensitivity.

Graphene, as a two dimensional material, is atomically thin and features high carrier mobility,[5] high mechanical strength[6,7] and piezoresistive electromechanical transduction.[8,9] At the same time, graphene technology is maturing and generally compatible with silicon semiconductor fabrication lines.[10] Therefore, graphene is an extremely interesting functional material for ultra-small nanoelectromechanical system (NEMS) devices.[8,11–15] Suspended atomically thin graphene structures that include doubly-clamped graphene beams, fully clamped graphene membranes and graphene cantilevers have been extensively studied and have been utilized in electromechanical resonators,[16–20] various types of pressure sensors,[9,21–28] strain sensors,[29,30] high responsivity photodetectors,[31] NEMS switches,[32] earphones,[33] loudspeakers,[34] microphones[35,36] and other NEMS devices.[7,37–42] NEMS accelerometers and gyroscopes typically require masses that are attached to suspended membranes, beams or cantilevers. However, realizing suspended graphene with large attached proof masses remians challenging. While there are reports of atomically thin graphene membranes with attached masses,[43–45] these masses are limited to extremely small and thin deposited coatings.

Recently, graphene ribbons with suspended silicon proof masses have been utilized as piezoresistive electromechanical transducers in NEMS accelerometers with direct electrical readout.[46] In these devices, an acceleration force acting on the proof mass caused a deflection of the mass. This in turn caused a change of the strain in the suspended graphene ribbons, resulting in a change of the electrical resistances of the graphene ribbons because of the piezoresistivity of graphene. While these devices were very small and sensitive, they had limitations such as relatively weak robustness, relatively low manufacturing yield, and part of them that had narrow ribbons did not survive high measurement currents. In the present paper, we report on piezoresistive NEMS accelerometer devices that are based on fully-clamped graphene membranes with suspended silicon proof masses, featuring several advantages over the previously demonstrated graphene ribbon devices, including improved mechanical robustness, suspension of larger silicon proof masses, high manufacturing yields of up to ~80%, and the capability to withstand higher measurement currents.

To demonstrate the utility of suspended graphene membranes with attached silicon proof masses as piezoresistive NEMS accelerometers, we have realized several device variations with different membrane and proof mass dimensions, all consisting of a suspended double-layer graphene membrane that is clamped at its entire circumference and that contains an attached silicon proof mass at its centre (**Figure 1**a). The working principle of our devices is based on the displacement of the proof mass that is caused by a force from external acceleration acting on the proof mass. The proof mass displacement results in a change of the strain in the suspended sections of the graphene membrane, thereby changing the resistances of the corresponding suspended sections of graphene membrane due to the piezoresistivity of graphene (**Figure 1**d1). The fabrication of our graphene NEMS devices is depicted in Figure 1b, with details of device fabrication described in the **Methods**. Briefly, chemical vapor deposition (CVD)[47] graphene was transferred from the original copper substrate to the silicon-on-insulator (SOI) device substrate, thereby suspending graphene over the 16.4 μm deep through-etched trenches in the oxidized silicon device layer of the SOI device substrate (**Figure 1**b1 and **Supporting Information** Figure S1 and S2). The silicon handle layer of the SOI substrate was removed in the areas below the trenches in the silicon device layer by deep reactive ion etching (DRIE), prior to graphene transfer and patterning (**Figure 1**b2). The graphene was then lithographically patterned using $O_2$ plasma etching. Thus, the proof masses defined in the silicon device layer of the SOI substrate were sandwiched between the graphene membranes and the buried $SiO_2$ (BOX) layer of the SOI substrate (**Figure 1**b3). In a final step, the BOX layer was sacrificially removed by dry plasma etching, followed by vapor HF etching to release the silicon proof masses and suspend them on the graphene membranes (**Figure 1**b4). After device fabrication, the devices were placed in a ceramic chip package and wire-bonded. **Figure 1**c shows photographs of two packaged chips (**Figure 1**c1 and c2), scanning electron microscope (SEM) images of two wire-bonded devices of the two packaged chips (**Figure 1**c3 and c4) and close-up images of the sensing regions of the corresponding devices (**Figure 1**c5 and c6). The two devices shown in **Figure 1**c5 and c6 represent two basic device designs with different geometric shapes of the graphene patch. The device shown in **Figure 1**c5 had a narrow graphene patch (we call it "type a" for short), while the device shown in **Figure 1**c6 had a wide graphene patch (we call it "type b" for short). The poof masses of all fabricated devices were 16.4 μm thick, including a 15 μm thick silicon layer and a 1.4 μm thick SiO$_2$ layer, and all proof masses had a quadratic shape with different side lengths ranging from 10-100 μm. The proof mass size 100 μm × 100 μm × 16.4 μm was the largest proof mass that we attempted to realize. The trench widths, i.e. the dimension defining the length of the freely suspended sections of the graphene membranes ranged from 2-4 μm, depending on the device dimensions.

Due to the complex series-parallel resistance connection in fully-clamped graphene membranes, we developed a resistance model to analyse the resistance change of the suspended sections of the graphene membranes. A schematic of the equivalent resistance model of our devices is shown in **Figure 1**d2, where R is the measured overall resistance of the graphene device, $R_{1l}$ is the resistance of the non-suspended graphene section between the left electrode and the left trench, $R_{1r}$ is the resistance of the non-suspended graphene section between right electrode and right trench, $R_{3t}$ is the resistance of the non-suspended graphene section above the top trench, $R_{3b}$ is the resistance of the non-suspended graphene section below the bottom trench, $R_{5l}$ is the resistance of the section of the suspended graphene over the left trench, $R_{5r}$ is the resistance of the section of the suspended graphene over the right trench, $R_{7t}$ is the resistance of the section of the suspended graphene over the top trench and $R_{7b}$ is the resistance of the section of the suspended graphene over the bottom trench, and $R_8$ is the resistance of the section of the non-suspended graphene on top of the surface of the proof mass. The top, bottom, left and right trenches of a device are depicted in **Figure 1**c5. Assuming ideal symmetry of the device (**Figure 1**d2), $R_{1l} = R_{1r} = R_1$, $R_{3t} = R_{3b} = R_3$, and $R_{5l} = R_{5r} = R_5$, $R_{7t} = R_{7b} = R_7$. The sheet resistance of the double-layer graphene patch is defined as $R_S$, and according to the square shape of the graphene areas that define $R_4$ and $R_8$, it follows that $R_4 = R_8 = R_S$. In this configuration, the variable resistances of the suspended graphene sections are $R_5$ and $R_7$, while $R_1$, $R_3$ and $R_8$ are assumed to be constant. For a measurement current I that is applied to a graphene device (**Figure 1**d2), the following equations can be derived based on the equivalent resistance model (see equation S1-S9 in **Supporting Information**).

$$R = R_{1l} + R_{1r} + R_2 = R_{1l} + R_{1r} + \frac{1}{\frac{1}{R_{3t}}+\frac{1}{R_4}+\frac{1}{R_{3b}}} \tag{1}$$

$$R_4 = R_{5l} + R_6 + R_{5r} \tag{2}$$

$$R_6 = \frac{1}{\frac{1}{R_{7t}}+\frac{1}{R_8}+\frac{1}{R_{7b}}} \tag{3}$$

The change of the overall resistance R of the graphene device as a result of a change in strain of the suspended graphene sections caused by a deflection of the proof-mass (ΔR) can thus be approximated by

$$\Delta R \approx \frac{2R_3^2 \times \Delta R_5}{(R_3+2R_4)^2} \tag{4}$$

From equation (4) as well as equation (S3) in the **Supporting Information** it can be seen that the influence of the resistance change in the suspended graphene sections across the top and bottom trenches ($\Delta R_{7t}$ and $\Delta R_{7b}$) on the total resistance change of a graphene device (ΔR) can be neglected, while the resistance changes of the suspended graphene sections over the left and right trenches ($\Delta R_{5l}$ and $\Delta R_{5r}$) are the dominant resistances contributing to the total resistance change of the graphene device (ΔR). Therefore, in the following text we define the resistances and resistance change of the suspended graphene sections over either the left or the right trench as $R_{SB}$ and $\Delta R_{SB}$, respectively, that is $R_{SB} = R_5 = R_{5l} = R_{5r}$ and $\Delta R_{SB} = \Delta R_5 = \Delta R_{5l} = \Delta R_{5r}$. Equation (4) describes the numerical relationship between the resistance change of the suspended graphene sections over the left and right trenches ($\Delta R_5$) and the total resistance change of a graphene device ($\Delta R$) and indicates that $\Delta R < 2\,\Delta R_5$. Thus, the resistance change of the suspended graphene sections over the left or right trenches ($\Delta R_{SB} = \Delta R_5 = \Delta R_{5l} = \Delta R_{5r}$) can be extracted according to the corresponding overall resistance change of the graphene device ($\Delta R$).

To evaluate the viability of our devices for detecting accelerations we used an air-bearing shaker (PCB 396C11, The Modal Shop) with a built-in high-precision reference accelerometer to expose our devices to accelerations at defined frequencies. In all experiments, the devices were placed with the sensitive axis in the direction of the earth gravitation, providing a 1 g acceleration bias. The packaged graphene NEMS devices were connected to signal read-out circuitry. The total resistance of a graphene device was the input resistance in the measurement circuit and the corresponding voltage output signal from the amplifier circuit (amplification factor of approximately 500) was recorded by a dynamic signal analyser (HP 35670A, Agilent Technologies Inc.) (**Methods and Supporting Information Figure S6**). The spectra of the amplified output voltages (amplification factor of approximately 500) from seven accelerometer devices (devices a1 to a7) and one reference device (reference) exposed to an acceleration of 1 g at a frequency of 160 Hz are shown in **Figure 2**a-h. In these measurements a bias supply current of 100 µA was supplied to the graphene device. The reference device consisted of a graphene patch that was similar to the graphene patches used in the accelerometer devices, but without etched trenches in the underlying substrate surface. Thus, no change in strain was introduced in the graphene patch of the reference device when exposed to accelerations. The output signals of devices a1 to a7 correlated with the acceleration, while no visible signal was observed in the output of the reference device. This confirmed that it was indeed the displacement of the proof mass of the accelerometer devices that caused the measured resistance change of the graphene and that there were no significant parasitic noise signals picked up by the graphene devices. SEM images of devices a1 to a7 and the reference device are shown in **Figure 2**i, and detailed information about the dimensions of each device can be found in Table S1 in the **Supporting Information**. As can be seen in **Figure 2**i, most of the devices contain one or more defects such as small holes in the suspended graphene membranes. In-plane tension, shear and compression of the suspended graphene, or occasional tears occurring at mechanically weak grain boundaries of CVD graphene are some possible causes for the holes, which were difficult to avoid during devices fabrication. Evaporation of water that could be encapsulated in the trenches after graphene transfer is another possible cause for rupture of the graphene membranes. In order to systematically compare the characteristics of devices a1-a7, they were exposed to an acceleration of 1 g at a frequency of 160Hz. The output voltage (U) and the overall measured resistance changes ($\Delta R$) of devices a1-a7 are shown in **Figure 3**a, while the resulting calculated resistance changes of the suspended parts of the graphene membranes over the left or right trench ($\Delta R_{SB} = \Delta R_5$) and the relative resistance changes of the suspended parts of the graphene membranes ($\Delta R_{SB}/R_{SB}$) of devices a1-a7 are shown in **Figure 3**b. The absolute resistance change of the suspended graphene sections over the left or right trenches ($\Delta R_{SB}$) was extracted (see Table S1 in the **Supporting Information)** based on equation (S9) in the **Supporting Information**, assuming that all suspended graphene sections in the graphene devices are ideal (without any defects and/or holes). Assuming that there are no defects such as holes in the suspended graphene sections, the device geometry determines that for an identical trench width, devices that have a smaller proof mass also have smaller total areas of the suspended graphene sections as compared to devices with a large proof mass. Interestingly, for device a1 (proof mass dimension of 10 μm × 10 μm × 16.4 μm) and device a2 (proof mass dimension of 20 μm × 20 μm × 16.4 μm), the output voltage (U), the overall resistance change of the graphene devices (ΔR) and the resistance change of the suspended graphene sections over the left and right trenches (ΔR$_{SB}$) were relatively higher compared to those of device a3 (proof mass dimension of 40 µm × 40 µm × 16.4 µm), device a4 (proof mass dimension of 40 µm × 40 µm × 16.4 µm) and device a6 (proof mass dimension of 50 µm × 50 µm × 16.4 µm), all devices having 3 µm wide trenches. However, all these devices featured similar relative resistance changes of the suspended graphene sections (ΔR$_{SB}$/R$_{SB}$). This indicates that in these devices the changes in the strain levels in the suspended graphene sections were of the same order of magnitude, although it should be noted that the relative resistance changes of the suspended graphene sections have a complex dependency of the trench width and the side length of the proof mass that determines both, the mass of the proof mass and trench length. In addition, the built-in stress in suspended graphene sections, and the number and position of the holes in the suspended graphene sections will influence the relative resistance changes of the suspended graphene sections. For devices with identical proof mass dimension (50 µm × 50 µm × 16.4 µm) and wider trenches such as device a5 (trench width of 4 µm), the output voltage (U), the overall resistance change of the device (ΔR), the resistance change of the suspended graphene sections over the left or right trenches (ΔR$_{SB}$) and the relative resistance change of the suspended graphene sections (ΔR$_{SB}$/R$_{SB}$) were higher compared to those devices with a narrower trench such as device a6 (trench width of 3 µm) and device a7 (trench width of 2 µm) (**Figure 3**a and b). In order to highlight the influence of the trench width on the output signal of a graphene device, a comparison of the output voltages of devices a5, a6 and a7 when exposed to an acceleration of 1 g at the frequency of 160 Hz are shown in **Figure 3**c. In this comparison of devices with identical proof mass dimensions, the devices with wider trenches featured larger output signals. A similar comparison of the influence of the trench length on the output signal is not easily possible because changing the trench length directly also changes the size and mass of the proof mass, thereby altering the overall device characteristics.

To demonstrate the expected linear relation between the measurement current and the output voltage of our graphene devices, device a6 (proof mass dimensions of 50 µm × 50 µm × 16.4 µm, trench width of 3 µm, one small hole) and device a8 (proof mass dimensions of 20 µm × 20 µm × 16.4 µm, trench width of 3 µm, several larger holes) were chosen for measurements at 1 g acceleration at a frequency of 160 Hz using different bias currents of 45 µA and 100 µA. The output voltages of both devices increased approximately proportionally with the measurement current, confirming that the measurement circuit worked as intended, even for the device that contain several holes in the graphene membrane (**Figure 4**a).

In order to analyse the influence of the number and dimensions of the holes in the suspended graphene sections on the output signals of the graphene devices, we compared measurements of devices a7, a9 and a10 when exposed to an acceleration at 1g at a frequency of 160 Hz. Devices a7, a9 and a10 had identical proof mass dimensions of 50 µm × 50 µm × 16.4 µm and trench widths of 2 µm, but different numbers and dimensions of holes in the graphene membranes, as shown in **Figure 2**i and **Figure 4**e. The comparison of the output voltage of devices a7, a9 and a10 when exposed to an acceleration of 1 g at a frequency of 160 Hz and a supply current of 100 µA was shown in **Figure 4**b. The results show that more holes in the graphene membrane and/or holes with larger dimension resulted in increased resistance changes of the devices. One possible explanation is that more holes or holes with larger dimensions result in reduced membrane stiffness and increased strain in the suspended graphene sections. It should be noted that no additional holes or visible defects were generated during the acceleration measurements (0-2g), which was verified by SEM imaging of the graphene devices before and after the acceleration measurements.

In order to verify the stability and repeatability of the output signal of our graphene devices, acceleration measurements of devices a11 and a12 were performed at different times (time span: 14 days). These devices had identical proof mass dimensions of 15 μm × 15 μm × 16.4 μm and an identical trench width of 4 μm but different number of defects in the suspended graphene sections (device a11 had one hole while device a12 had three holes as shown in **Figure 4**e). Device a11 exhibited good repeatability of the output signals at applied accelerations (1g, 1.5g and 2g), as shown in **Figure 4**c while device a12 presented a much lower repeatability of the output voltages at applied accelerations (1g, and 2g), as shown in **Figure 4**d. This indicates that the number of defects in the suspended graphene sections can influence the repeatability and stability of the output signal of a device. One or two small holes did not significantly influence the stability of the output signal of the graphene device (e.g. device a11) while several larger holes had a significant impact on the stability of the output signal of the device (e.g. device a12). This is expected as it suggests that the better the quality and integrity of the suspended graphene membranes, the better the repeatability of the output signals of the device.

In order to compare the influence of the geometric design of the graphene patch of a device on the resulting output signal, we performed corresponding acceleration measurements of "type b" devices that feature wider graphene patches than the "type a" devices (**Figure 5**i). Device b1 (proof mass dimensions of 10 μm × 10 μm × 16.4 μm and trench width of 4 μm), and device b2 (proof mass dimensions of 40 μm × 40 μm × 16.4 μm and trench width of 4 μm) are two examples of "type b" device designs. The spectra of the amplified output voltages of devices b1 and b2 exposed to an acceleration of 1 g at a frequency of 160 Hz are shown in **Figure 5**a and b. The corresponding output voltages (U) and the overall resistance changes ($\Delta$R) of devices b1 and b2 are shown in **Figure 5**c, and the resistance changes of the suspended graphene sections ($\Delta R_{SB}$) over the left or right trenches and the corresponding relative resistance changes ($\Delta R_{SB}/R_{SB}$) of devices b1 and b2 are shown in **Figure 5**d. These results illustrate that "type a" device designs (devices a1-a7) featured higher output signals than "type b" device designs (devices b1 and b2) (**Figure 5**e and f). Based on our resistance model of the devices, this result is expected because the fixed resistances that are coupled in parallel to the variable resistances are significantly larger in the "type a" device designs (**Figure 1**d2). **Figure 5**e shows the measured output voltages of devices a1 and b1 when exposed to an acceleration of 1 g at a frequency of 160 Hz. Devices a1 and b1 had identical proof mass dimensions of 10 μm × 10 μm × 16.4 μm. Although the trench width of device b1 (4 μm) was larger than the trench width of device a1 (3 μm), the output voltage of device a1 was significantly larger than that of device b1. A comparison of the relative resistance changes of the suspended graphene sections ($\Delta R_{SB}/R_{SB}$) of devices a1 and b1 is shown in **Figure 5**f. The measured values of $\Delta R_{SB}/R_{SB}$ of devices a1 and b1 were comparable, which was expected as they had identical proof mass dimensions and similar trench widths.

In summary, we have demonstrated that proof masses with dimensions of at least up to 100 μm × 100 μm × 16.4 μm can be suspended on membranes made of double-layer graphene. These devices can be utilized as piezoresistive NEMS accelerometer devices, where there exist trade-offs between device miniaturization (proof mass dimensions and trench widths), resulting output signal, device robustness and measurable acceleration range. AFM indentation experiments demonstrated that a device with a 3 μm wide trench and a proof mass size of 10 μm × 10 μm × 16.4 μm was able to withstand an indentation force at the proof mass center of up to 4070 nN, and ruptured at an indentation force of 5051 nN (**Supporting Information** Figure S7 and Table S2). For reference, this can be compared to the resulting force of about 0.039 nN resulting from an acceleration of 1 g on a proof mass of the size of 10 µm × 10 µm × 16.4 µm. We hypothesize that device designs with annular graphene membranes and proof masses may feature even higher mechanical robustness by avoiding corners that are prone to stress concentrations. In addition, annular device design would feature uniform strain distribution in all suspended graphene sections. Furthermore, it should be noted that the adhesion between a graphene sheet and a SiO$_2$ surface is known to be ultra-strong.[48] Thus, for any realistic acceleration force we do in our devices not expect delamination of the graphene from the substrate or the proof mass. Compared to state-of-the-art silicon MEMS accelerometers, our NEMS accelerometer structures feature at least three orders of magnitude smaller proof masses,[49-54] while still providing useful output signals. Consequently, our devices occupy at least two orders of magnitude small die areas, thereby demonstrating the huge potential of this approach for device scaling and cost reduction. In addition, there is further potential for reducing dimensions in our graphene devices by minimizing the electrical contact areas and by using state-of-the-art device packaging strategies.[55] Compared to our recent study on devices based on graphene ribbons with suspended silicon proof masses,[46] the fully-clamped graphene membrane devices feature improved fabrication yield, improved robustness, a potentially longer life span, the potential to measure higher accelerations, they can suspend larger proof masses and have the capability to withstand higher measurement currents. We have applied measurement currents of up to 200 µA to our graphene membrane devices and we did not observe failure of any of the devices caused by the measurement currents. The capability to withstand higher measurement currents of devices made of fully clamped graphene membranes is presumably a result of lower current densities for a given bias voltage in the graphene membranes compared to the graphene ribbons. However, the signal response of the devices with fully-clamped graphene membranes is generally lower than that of devices with doubly-clamped graphene ribbons for devices with identical proof masses and trench width. One reason for this is that for exposure to a given acceleration, significantly larger strain changes will be generated in narrow graphene ribbons with a proof mass as compared to suspending the identical proof mass on a graphene membrane that is fully clamped at its circumference. Another important reason is that the unavoidable parallel connected fixed graphene resistances in devices with fully-clamped graphene membranes significantly reduces the overall resistance response of these designs as compared to the simple series connection of the variable graphene resistances in devices based on graphene ribbons with a suspended proof mass.[46] While the piezoresistivity of graphene is most likely the dominant transduction mechanism in our devices, it is possible that changes of the interlayer interaction in double-layer graphene, such as sliding of individual graphene layers with respect to each other, can contribute to the measured resistance changes in our devices.[56] However, this effect is not likely to cause significant resistance changes in our graphene membranes because the required forces and membrane displacements indicated by Benameur *et al.*[56] to cause measurable resistance changes by this effect are much larger than the forces occurring in our acceleration.

**Conclusions**

The proposed fully-clamped graphene membranes with attached silicon proof masses for use in piezoresistive NEMS accelerometers feature excellent characteristics, including improved robustness, improved manufacturing yield, longer life time, the capability to withstand higher measurement currents and the potential to suspend larger proof masses than similar structures previously reported. These characteristics will contribute to bringing ultra-miniaturized piezoresistive graphene NEMS closer towards deployment in potential applications, spanning several important scientific and technological areas, such as IoT, biomedical implants, nanoscale robotics, and wearable electronics.

**Methods**

*Device fabrication*

Devices were fabricated from a SOI wafer in which the silicon device layer was 15 μm thick, the BOX layer was 2 μm thick and the handle substrate was 400 μm thick (**Figure 1**a). First, a 1.4 μm thick layer of $SiO_2$ was thermally grown on both the front and the backside of the SOI wafer (**Figure 1**a). Next, a photoresist layer was spin-coated on the $SiO_2$ surface of the silicon device layer and patterned to define the metal electrodes. The pattern was transferred into the 1.4 μm thick $SiO_2$ layer by etching 300 nm deep cavities using reactive ion etching (RIE). The cavities were filled with a 50 nm thick layer of titanium (Ti) followed by a 270 nm thick layer of gold (Au) using metal evaporation. The photoresist layer was removed in a lift-off process by wet etching, leaving the patterned Au electrodes protrude by about 20 nm above the $SiO_2$ surface. A new photoresist layer was spin-coated on the $SiO_2$ surface and lithographically patterned for defining the trenches surrounding the proof masses. RIE was used to etch through the 1.4 μm thick $SiO_2$ layer and DRIE was used to etch through the 15 μm thick silicon device layer (**Figure 1**b1). After the DRIE, photoresist residues were removed by $O_2$ plasma etching. Next, a photoresist layer was spin-coated and patterned on the $SiO_2$ surface of the backside of the SOI wafer, defining squares with dimensions of 150 μm × 150 μm placed in the same areas that define the proof masses in the silicon device layer. Using this mask, the 1.4 μm thick $SiO_2$ layer was etched by RIE. Next, the handle substrate was etched by DRIE until reaching the BOX, using both the photoresist and the $SiO_2$ as masking layers, thereby forming the channel to release the proof mass in a later process step. Photoresist residues were removed by $O_2$ plasma etching, which finalized the pre-processing of the SOI device substrate (**Figure 1**b2). The device substrate was then diced in 8 mm × 8 mm large chips, each containing 64 devices.

Commercially available CVD single-layer graphene on copper foil (Graphenea, Spain) was used in this work. Double-layer graphene was obtained by transferring a single-layer graphene to another single-layer graphene on a copper foil. Therefore, a poly (methyl methacrylate) (PMMA) solution (AR-P 649.04, ALLRESIST, Germany) was spin-coated on the front-side of the first graphene/copper foils at 500 rpm for 5 s and at 1800 rpm for 30 s and then baked for 5 minutes at 85°C on a hot plate to evaporate the solvent and cure the PMMA, resulting in a PMMA film thickness of about 200 nm. Carbon residues on the backside of the copper foil were removed using $O_2$ plasma etching at low power (50-80 W). In order to release the graphene/PMMA stack from the copper, the foil was placed onto the surface of an iron chloride ($FeCl_3$) solution with the graphene side facing away from the liquid, resulting in wet etching of the copper. After 2 hours, the PMMA/graphene stack was transferred onto the surface of deionized (DI) water, then diluted HCl solution and, back to DI water for cleaning, removing the iron (III) residues and removing chloride residues, respectively. A silicon wafer was used for handling and picking up the PMMA/graphene stack from the liquids. A second graphene on copper foil was used and the PMMA/graphene stack floating on the DI water was transferred to the second graphene on copper foil and subsequently put on a hotplate at 45°C to increase the adhesion between the two graphene layers. Carbon residues on the backside of the copper foil were removed using $O_2$ plasma etching. Again, the same processes were performed to remove the copper substrate from the double-layer graphene and transfer the final PMMA/double-layer graphene stack to the pre-processed SOI device substrate. The SOI device substrate was then baked for ~ 10 minutes at ~ 45°C in order to dry it and to increase the bond strength between the double-layer graphene and the $SiO_2$ surface. Next, the SOI device substrate was placed in acetone to remove the PMMA and subsequently in isoproponal to remove acetone residues. It should be noted that even after an exhaustive rinse with organic solvents, there are some PMMA residues (long-chain molecules) remaining on the graphene (Figure S1 in **Supporting Information**) due to the strong dipole interactions between PMMA and the chemical groups on graphene.[57] After the graphene transfer, a photoresist layer was spin-coated on the graphene at 1000 rpm for 5 seconds and 4000 rpm for 60 seconds and then baked on a hotplate for 30-60 seconds at 90°C. Optical lithography and photoresist development were done using a standard developer for 15 seconds and DI water for 10 seconds for rinsing, and then the SOI substrate was dried in air. Next, the graphene was etched by $O_2$ plasma at 50 W for 120 seconds to define the outline of the patches with graphene membranes. In order to remove the photoresist residues, the device substrate was placed in acetone for 20 minutes and then in isopropanol for 5 minutes, followed by baking at 45°C for 10 minutes on a hotplate (**Figure 1**b3).

In order to release the proof masses and suspend them on the double-layer graphene membranes, the BOX layer (2 μm thick $SiO_2$) was partly etched from the backside of the SOI substrate by RIE, followed by vapor HF etch to remove the remaining $SiO_2$ layer (**Figure 1**b4). This two-step etching process was employed to minimize the risk of damaging the graphene. For etching the BOX layer, the device substrate was attached to a 100 mm diameter silicon carrier wafer and the 4 edges on the sides of the device substrate were sealed with a tape. Then RIE etching was employed to etch approximately 1.9 μm of the BOX layer, leaving a ~100 nm thin BOX layer that was suspending the silicon proof masses. Vapor HF was then used to etch the remaining BOX layer using a custom-build vapor HF etching setup. 25 % of HF was used in the vapor HF chamber and the temperature was set to 40°C. The etching of the ~100 nm thick BOX layer typically took 5-10 minutes. Once the device fabrication was complete, the chips were glued in a ceramic chip package with an open cavity. Next, gold wire bonding was used to connect the electrode pads on the device substrate to the bond pads in the chip package as shown in **Figure 1**c. Our fabrication process resulted in a fabrication yield of about 80% of fully-clamped double-layer graphene membranes with attached silicon proof masses that were electrically active. Our graphene devices (both "type a" and "type b" devices) were functioning after elaborate acceleration measurements and most of the time even after exposure to shocks during device handling. Only in very rare cases devices were damaged when exposed to excessive shocks during handling. This illustrates the increased robustness, fracture toughness[58] and life-span of the devices compared to devices based on doubly-clamped graphene ribbons that were in many cases damaged when exposed to shocks during handling. It should be noted that our fabrication yield was very low (on the order of 1%) when we attempted to realize membranes made of single-layer graphene with attached silicon proof masses, and it was very difficult to manually handle these devices without breaking them. Although CVD graphene intrinsically comprises grains and grain boundaries that might influence the mechanical strength and fracture toughness of CVD graphene, our experimental results illustrate that for a suspended structure made of CVD graphene, the addition of a second layer of CVD graphene on top of a first CVD graphene layer disproportionally increases the fracture toughness of the resulting structure. This is consistent with literature reports showing that suspended membranes made of double-layer CVD graphene have a better overall mechanical resilience compared to suspended membranes made of single-layer CVD graphene.[58] However, if single-layer graphene devices can be successfully fabricated they potentially would feature higher output signals as compared to double-layer graphene devices due to the reduced stiffness of single-layer graphene membranes. We also hypothesize that tri-layer or multi-layer graphene membranes in our devices would further increase the fabrication yield but likely result in reduced output signals.

*Characterizations and measurements*

Optical microscopy, white light interferometry (Wyko NT9300, Veeco), and SEM imaging were used to observe and characterize the morphology of the devices during and after device fabrication (Figure S1-S4 in **Supporting Information**). SEM imaging was used to evaluate graphene membranes and suspended proof masses after they were released by sacrificial etching of the BOX layer (Figure S1 and S2 in **Supporting Information**). White light interferometry was used to detect $SiO_2$ residues inside the trenches to verify that the masses were fully released, and to measure the deflection of the silicon proof masses in relation to the substrate surface after release of the proof masses (Figure S3-S4 in **Supporting Information**). Raman spectrometry (alpha300 R, WITec) was used to verify the presence and quality of the double-layer graphene of a fabricated device (Figure S5 in **Supporting Information**). An atomic force microscope (AFM) (Dimension Icon, Bruker) with a cantilever (Olympus AC240TM) and an AFM tip (tip radius = 15 nm) was used to load defined forces at the center of a proof mass of a graphene device to measure the force versus proof mass displacement as well as the maximum force that the suspended graphene membrane can withstand without rupture (Figure S7 and Table S2 in **Supporting Information**). A probe-station connected to a parameter analyzer (Keithley SCS4200, Tektronix) was used for preliminary electrical characterization of graphene devices. A special box with a complete low-frequency electromagnetic shielding was designed to shield the graphene devices from mechanical and electrical noise interferences of the measurement system and the environment. Both the device package and the electronic measurement circuits were encapsulated in this box but were separated by ferromagnetic alloy inside the box to reduce crosstalk between the graphene devices and the measurement circuit. The air-bearing shaker and the dynamic signal analyser were controlled by a computer interface to apply a defined acceleration for the packaged devices and to read out the sensing signal in form of an output voltage, respectively. In all experiments we used an acceleration frequency of 160 Hz with a 1 g gravitation bias. The acceleration frequency of 160 Hz was used because on one hand, this frequency is sufficiently high to obtain relatively low 1/f-noise and is sufficiently close to 159.2 Hz, which is a commonly used frequency for accelerometer calibrators and is equivalent to a radian frequency of 1000 rad/s (equivalent to $2 \times \pi \times 159.2$)[59] and, on the other hand, it is well below the intrinsic resonance frequencies of the spring-mass systems of our devices. In addition, 160 Hz avoids the commonly known 50 Hz noise sources and its multiples. The measurement circuits consisted of a first-order high-pass filter with a cut off frequency of 0.079 Hz, and a preamplifier (LT1001OP, Linear Technology) with an amplification factor of approximately 500. The resistance of the graphene patch of a device was used as the input resistance of the first-order high-pass filter, supplied with an adjustable DC current. Unless stated differently, in all experiments a bias current of 100 µA was used and the measurements were performed in atmospheric conditions. When the proof mass was displaced, the suspended sections of the graphene membrane were strained and their resistances changed due to the piezoresistivity of graphene. The output voltage induced by the change of the graphene resistance was read by the first-order high-pass filter with a cut-off frequency of 0.079 Hz to filter any DC drift observed in the graphene resistors, amplified by the amplifier, read by a first-order high-pass filter with a cut-off frequency of 15 Hz and then recorded by a dynamic signal analyser (HP 35670A), and finally recorded through the computer interface (Figure S6 in **Supporting Information**). According to the output voltage U, the corresponding resistance change of graphene devices can be directly extracted by $\Delta R = U/(500 \times I)$. Before and after each measurement, a multimeter was used to measure the resistance of the graphene patch, in order to confirm that the graphene membrane with the suspended mass was intact before and after the measurements. In addition, optical microscopy and SEM imaging were used to confirm the mechanical integrity of the devices after the measurements. All characterizations and measurements were performed in air at atmospheric pressure.

**ASSOCIATED CONTENT**

**Supporting Information**

The Supporting Information is available free of charge on the ACS Publications website at http:// pubs.acs.org.

Supplementary equations: the equivalent resistance model of graphene devices (equation (S1-S9)). Supplementary figures: SEM images of a graphene device with high magnification (Figure S1); SEM characterization of the backside of fabricated graphene devices (Figure S2); demonstration of release of the proof masses using white light interferometry (Figure S3); measurement of static displacement of the released proof masses in relation to the substrate surface using white light interferometry (Figure S4); Raman spectroscopy of the double-layer graphene on a device (Figure S5); measurement circuit and experimental setup (Figure S6); force-displacement measurements of suspended graphene membranes with attached proof mass using AFM tip indentation (Figure S7); Supplementary tables: Graphene device dimensions and extracted resistances of the different graphene sections of the device (Table S1); Displacement of the suspended silicon proof mass at different applied AFM indentation forces (Table S2).

**AUTHOR INFORMATION**

**Corresponding Authors**

*E-mail: (X. F.) xuge@eecs.kth.se

*E-mail: (M.C.L.) lemme@amo.de

*E-mail: (F.N.) frank.niklaus@eecs.kth.se

**Author Contributions**

X.F., F.N., F.F., A.D.S., and M.C.L. conceived and designed the experiments. A.D.S., S.W., M.Ö. and M.C.L. contributed in the graphene transfer. S.W. contributed in Raman characterization. S.S. contributed in the packaging of all devices. F.F. designed the measurement circuits and contributed in acceleration measurements. X.F. performed all the experiments, including device fabrication, graphene transfer and pattern, various characterizations (optical microscopy, white light interferometry, SEM imaging, AFM indentation, and electrical characterization based on probe-station), acceleration measurements and wrote the manuscript. F.N. provided guidance in all the experiments and manuscript writing. X.F., F.F., F.N., and M.C.L. analyzed the experimental results. All authors discussed the results and commented on the manuscript.

**Notes**

The authors declare no competing financial interest.

**ACKNOWLEDGEMENTS**

This work was supported by the European Research Council through the Starting Grant M&M's (No. 277879) and InteGraDe (307311), the China Scholarship Council (CSC) through a scholarship grant, the Swedish Research Council (GEMS, 2015-05112), the German Federal Ministry for Education and Research project NanoGraM (BMBF, 03XP0006C), the German Research Foundation (DFG, 750 LE 2440/1-2) and the European Commission (Graphene Flagship, 785219). The authors thank Andreas C. Fischer, Henrik

Rödjegård and Guillermo Villanueva for helpful discussions, Cecilia Aronsson for support with the device processing and Mikael Bergqvist for assistance with the measurement setup.

(16) Bunch, J. S.; Zande, A. M. van der; Verbridge, S. S.; Frank, I. W.; Tanenbaum, D. M.; Parpia, J. M.; Craighead, H. G.; McEuen, P. L. Electromechanical Resonators from Graphene Sheets. *Science* **2007**, *315* (5811), 490–493. https://doi.org/10.1126/science.1136836.

(17) Miao, T.; Yeom, S.; Wang, P.; Standley, B.; Bockrath, M. Graphene Nanoelectromechanical Systems as Stochastic-Frequency Oscillators. *Nano Letters* **2014**, *14* (6), 2982–2987. https://doi.org/10.1021/nl403936a.

(18) Will, M.; Hamer, M.; Müller, M.; Noury, A.; Weber, P.; Bachtold, A.; Gorbachev, R. V.; Stampfer, C.; Güttinger, J. High Quality Factor Graphene-Based Two-Dimensional Heterostructure Mechanical Resonator. *Nano Lett.* **2017**, *17* (10), 5950–5955. https://doi.org/10.1021/acs.nanolett.7b01845.

(19) Zande, A. M. van der; Barton, R. A.; Alden, J. S.; Ruiz-Vargas, C. S.; Whitney, W. S.; Pham, P. H. Q.; Park, J.; Parpia, J. M.; Craighead, H. G.; McEuen, P. L. Large-Scale Arrays of Single-Layer Graphene Resonators. *Nano Lett.* **2010**, *10* (12), 4869–4873. https://doi.org/10.1021/nl102713c.

(20) Chen, C.; Rosenblatt, S.; Bolotin, K. I.; Kalb, W.; Kim, P.; Kymissis, I.; Stormer, H. L.; Heinz, T. F.; Hone, J. Performance of Monolayer Graphene Nanomechanical Resonators with Electrical Readout. *Nature Nanotechnology* **2009**, *4* (12), 861–867. https://doi.org/10.1038/nnano.2009.267.

(21) Smith, A. D.; Niklaus, F.; Paussa, A.; Vaziri, S.; Fischer, A. C.; Sterner, M.; Forsberg, F.; Delin, A.; Esseni, D.; Palestri, P.; et al. Electromechanical Piezoresistive Sensing in Suspended Graphene Membranes. *Nano Lett.* **2013**, *13* (7), 3237–3242. https://doi.org/10.1021/nl401352k.<007>

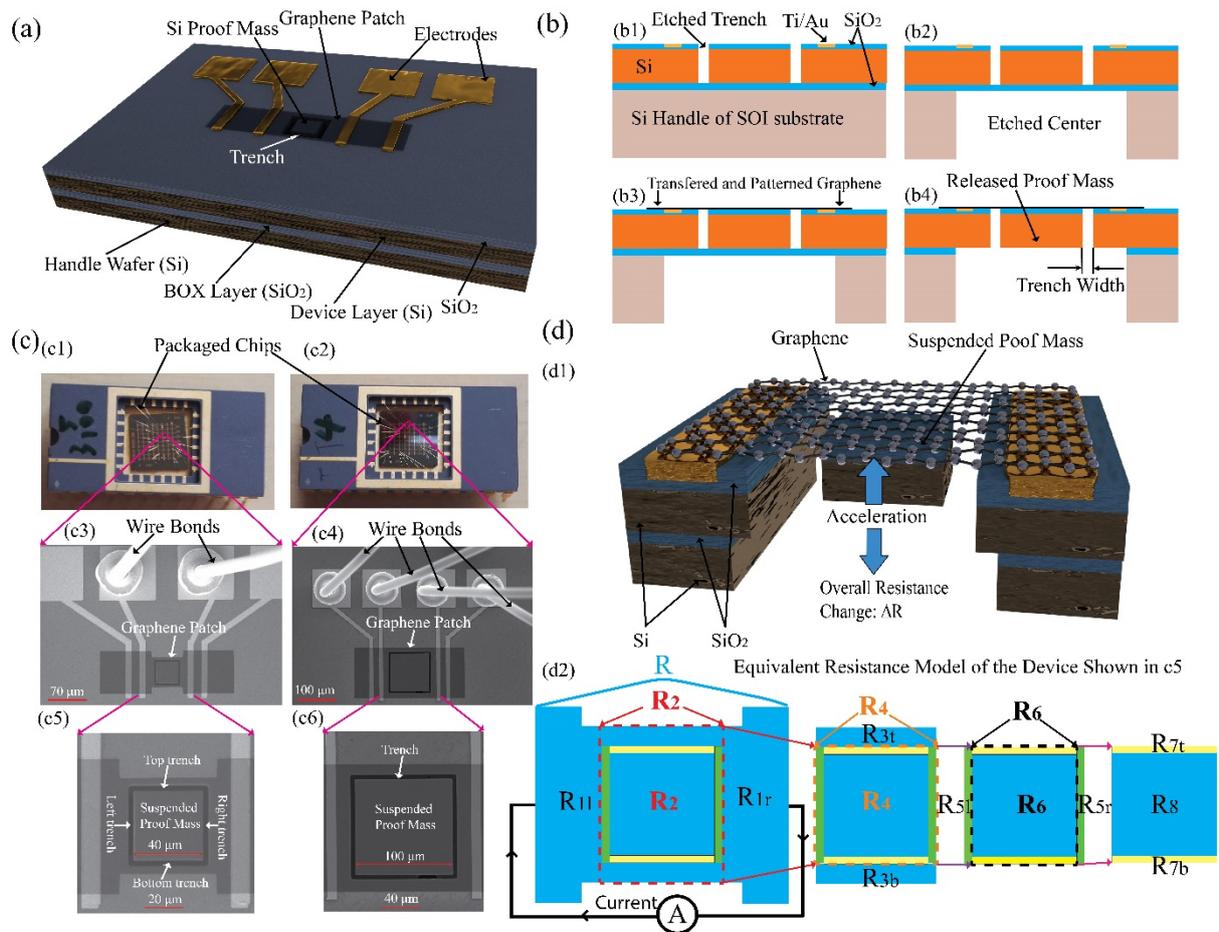

**Figure 1.** Suspended graphene membrane with attached silicon proof mass as piezoresistive NEMS accelerometers. (a) 3D illustration of the device design. (b) Schematic of device fabrication. b1, Ti/Au electrodes are embedded into a 1.4 μm thick SiO$_2$ layer of an SOI substrate. Trenches were etched in the 15 μm thick silicon device layer of the SOI substrate to form the silicon proof mass. b2, The 400 μm thick silicon handle layer of the SOI substrate was etched by DRIE in the areas below the proof masses, leaving the silicon proof masses suspended on the BOX layer. b3, Double-layer graphene was transferred to the SiO$_2$ surface of the SOI substrate and patterned by photoresist masking and O$_2$ plasma etching. b4, The silicon proof masses were released by sacrificially etching the BOX layer in a 2-step etching process, using first RIE etching followed by vapor HF etching. (c) Packaging and wire bonding of the devices. c1 and c2, photographs of two packaged and wire bonded dies with "type a" devices and "type b" devices, respectively. c3 and c5, SEM images of a "type a"

device from (c1) and the corresponding close-up. c4 and c6, SEM images of a "type b" device from (c2) and the corresponding close-up. The trench width in the device in c5 was 3 µm while the trench width in the device in c6 was 4 µm. The side lengths of the squared masses in the devices in c5 and c6 were 40 µm and 100 µm respectively. (d) Transduction principle and equivalent resistance model of the graphene devices. d1, Cross-sectional 3D illustration of a fully clamped graphene membrane with attached proof mass. The deflection of the proof mass and the resulting strain in the suspended graphene sections causes resistance changes in the suspended graphene due to the piezoresistivity of the graphene. d2, Equivalent resistance model of the graphene devices. The resistances of graphene areas on the SiO$_2$ surface are represented by $R_{1l}$ and $R_{1r}$ ($R_{1l} = R_{1r} = R_1$), $R_{3t}$ and $R_{3b}$ ($R_{3t} = R_{3b} = R_3$), and R$_8$ respectively. The resistances of the suspended graphene sections over the trenches are represented by $R_{5l}$ and $R_{5r}$ ($R_{5l} = R_{5r} = R_5$), and $R_{7t}$ and $R_{7b}$ respectively ($R_{7t} = R_{7b} = R_7$).

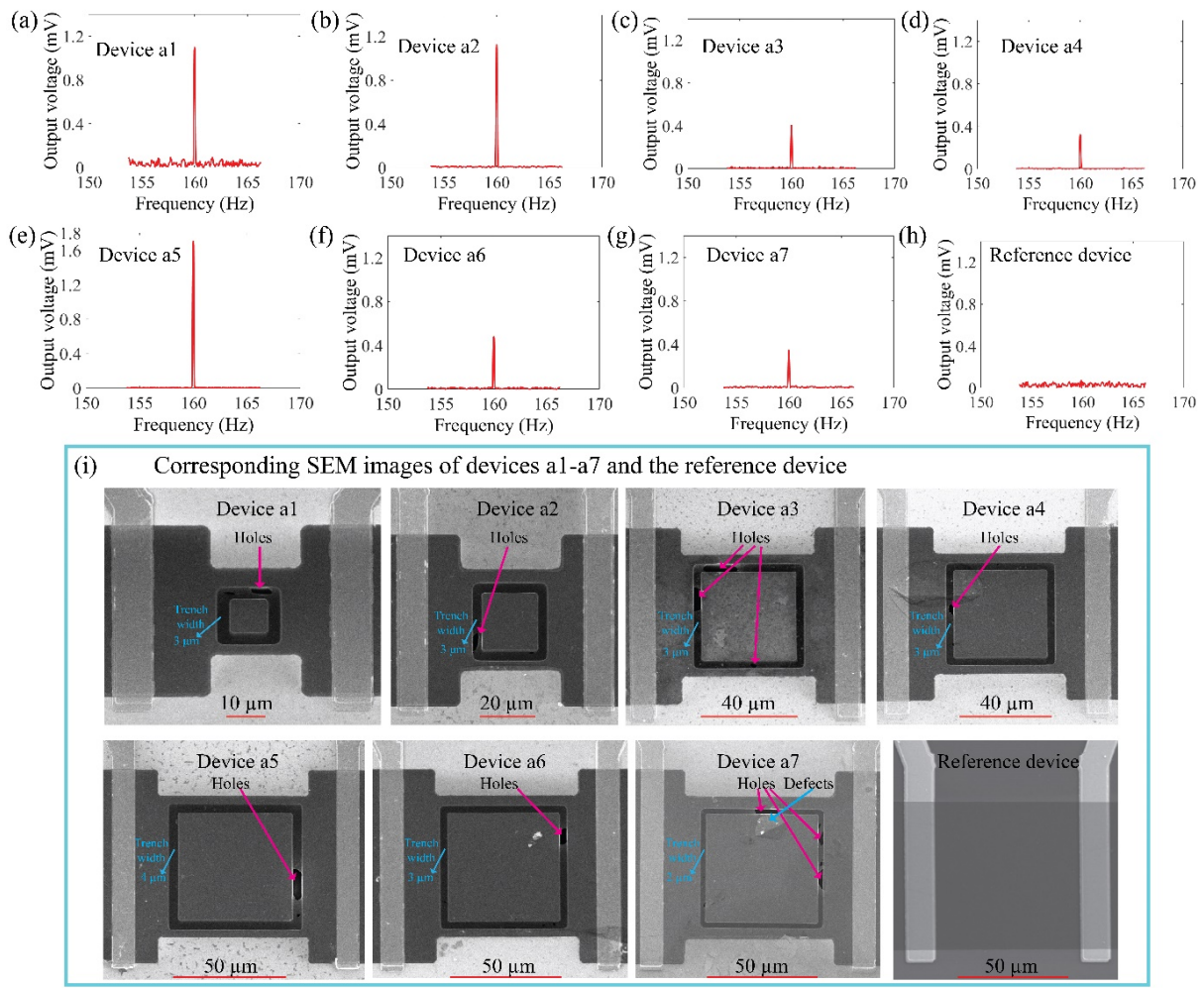

**Figure 2.** Measured spectra of the output voltages of graphene devices ("type a"). (a-h) Measured spectra of the output voltage of devices a1 to a7 and of the reference device when exposed to an acceleration of 1 g at a frequency of 160 Hz. In devices a1 to a7, the acceleration caused a deflection of the proof masses, and consequently the strain and resistance changed in the suspended graphene sections. This dependence on applied acceleration is not observed in the reference device. (i) Corresponding SEM images of devices a1 to a7 and the reference device. In all devices there are one or more small holes in the suspended graphene sections that were caused during devices fabrication, which could not be avoided.

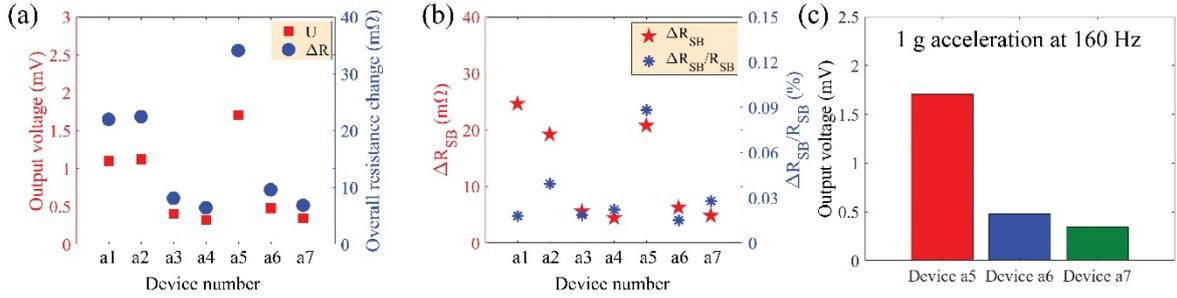

**Figure 3.** Comparison of characteristics of the output signals of devices a1 to a7. (a) Comparison of the measured output voltages and overall resistance changes of devices a1 to a7 when exposed to an applied acceleration of 1 g at a frequency of 160 Hz. (b) Comparison of the calculated resistance changes of the suspended graphene sections ($\Delta R_{SB}$) across either the left or the right trench and the corresponding relative resistance changes ($\Delta R_{SB}/R_{SB}$) of devices a1 to a7 when exposed to an applied acceleration of 1 g at a frequency of 160 Hz. (c) Comparison of the output voltages of devices a5, a6 and a7 when exposed to an applied acceleration of 1 g at a frequency of 160 Hz. Devices a5 to a7 have proof masses with identical dimensions (50 μm × 50 μm × 16.4 μm), but different trench widths (4 μm for device a5, 3 μm for device a6, and 2 μm for device a7).

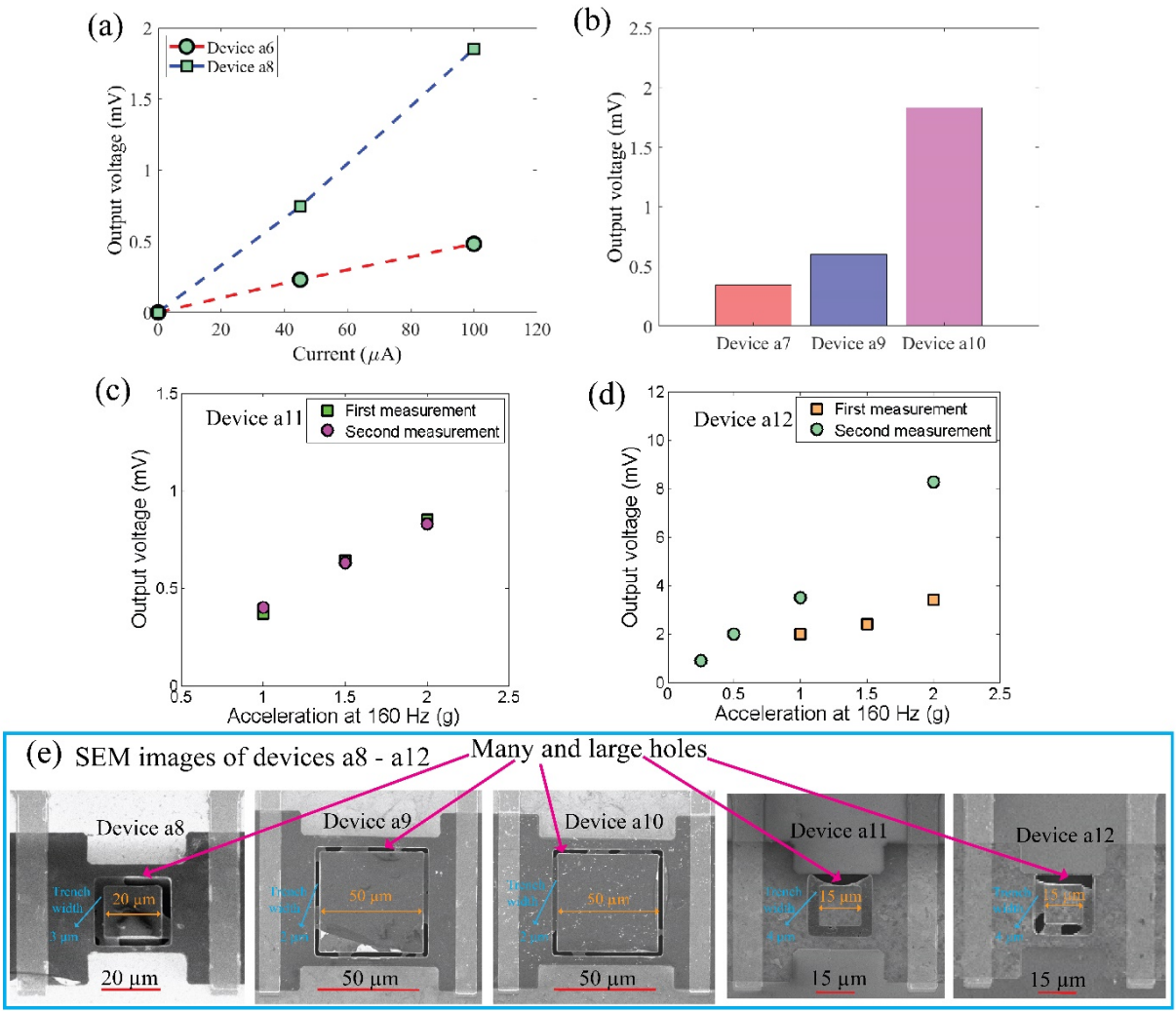

**Figure 4.** Control experiments of device output signals. (a) Measured output voltages versus bias currents of devices a6 and a8. The output voltage increased nearly proportionally with the current bias, confirming that the measurement setup works as intended. (b) Comparison of the output voltages of devices a7, a9 and a10 when exposed to an acceleration of 1 g with a frequency of 160 Hz. Devices a7, a9 and a10 had identical proof mass dimensions of 50 μm × 50 μm × 16.4 μm and identical trench widths of 2 μm, but different numbers and sizes of defects in the suspended graphene membranes. Device a7 had fewer and smaller holes and featured a relatively small output voltage while devices a9 and a10 had many and larger holes and featured higher output voltages. This illustrates that the number of holes in the suspended areas of the graphene influences the signal response of a fully-clamped graphene device. The large difference of the output voltages of devices a9 and a10 illustrates instabilities of devices

with a large number of defects in the suspended sections of the graphene membrane. (c) Comparison of the output voltages of device a11 when exposed to accelerations at a frequency of 160 Hz for different measurements on different days. The measurement results illustrate that device a11 with few defects in the graphene membrane featured good repeatability. (d) Comparison of the output voltages of device a12 when exposed to applied acceleration of 1 g at a frequency of 160 Hz for different measurements on different days. The measurement results illustrate that device a12 with a larger number of defects in the graphene membrane featured significantly lower measurement repeatability but increased output voltages as compared to device a11 with fewer defects. (e) SEM images of devices a8 to a12. Devices a8 to a12 were "type a" devices (device with narrow graphene patch). Device a11 had only one defect and has relatively stable device characteristics. Devices a8, a9, a10 and a12 had a large number of defects and featured comparably unstable output signals.

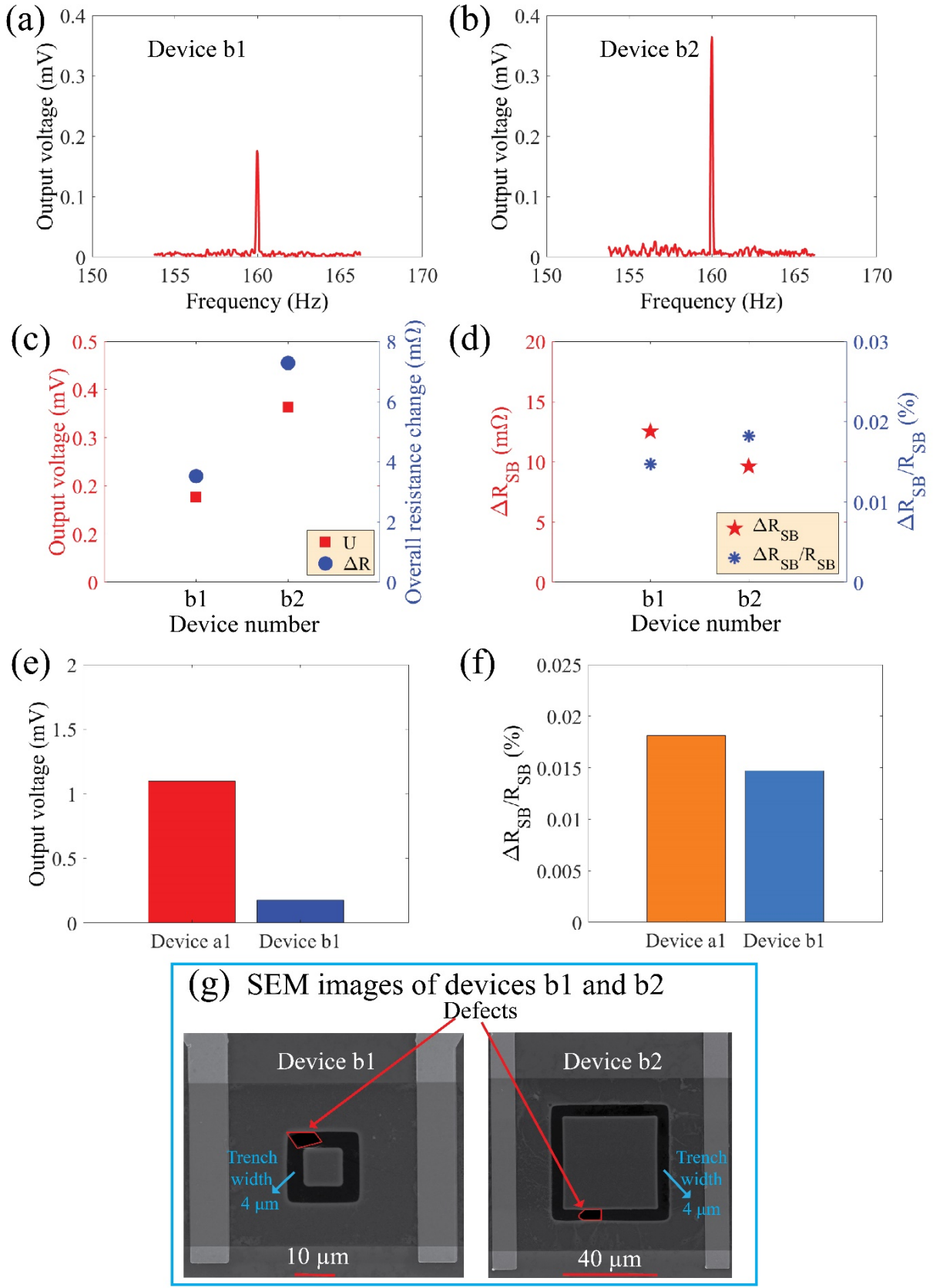

**Figure 5.** Measured output voltages of "type b" devices and comparison to "type a" devices. (a-b) Measured spectra of the output voltages of devices b1 and b2 when exposed to an applied acceleration of 1 g at a frequency of 160 Hz. (c) Comparison of the output voltages and overall resistance changes (ΔR) of devices b1 and b2 when exposed to an applied acceleration of 1 g at a frequency of 160 Hz. (d) Comparison of the calculated resistance change of the suspended graphene section ($\Delta R_{SB}$) in either the left or the right trench and corresponding relative resistance change ($\Delta R_{SB}/R_{SB}$) of devices b1 and b2 when exposed to an applied acceleration of 1 g with a frequency of 160 Hz. (e) Comparison of the output voltages of devices a1 and b1 when exposed to an acceleration of 1 g at a frequency of 160 Hz. Devices a1 and b1 had identical proof mass dimensions of 10 μm × 10 μm × 16.4 μm and similar trench widths (3 μm for device a1 and 4 μm for device b1), but device a1 was a "type a" device design (device with a narrow graphene patch) while device b1 was a "type b" device design (device with a wide graphene patch). The results of (e) illustrate that "type a" designs typically featured higher signal responses than "type b" designs under identical conditions. (f) Comparison of the calculated relative resistance changes of the suspended graphene section ($\Delta R_{SB}/R_{SB}$) of devices a1 and b1, when exposed to an acceleration of 1 g at a frequency of 160 Hz, respectively. (g) SEM images of devices b1 and b2 ("type b").